\documentclass[10pt,twocolumn,twoside]{IEEEtran}
\usepackage{latexsym}
\usepackage{amsmath}
\usepackage{amssymb}
\usepackage{epsfig}
\usepackage{cite}
\usepackage{algorithmic}
\usepackage{overpic}
\usepackage{multirow}
\usepackage[table]{xcolor}
\usepackage{graphicx}
\usepackage{colortbl,array}
\usepackage{multirow,bigdelim}
\usepackage{graphicx}
\usepackage{subfigure}
\usepackage{booktabs}
\usepackage{booktabs,amsmath}
\usepackage{fancyhdr, graphicx, amsmath, amssymb}
\usepackage{nomencl}
\newtheorem{thm}{Theorem}

\newtheorem{rem}{Remark}

\ifCLASSINFOpdf
\else
\fi

\begin{document}

\title{Dynamics of Public Opinion Evolution with Asymmetric Cognitive Bias}

\author{Yanbing~Mao, Naira~Hovakimyan, and Tarek~Abdelzaher
\thanks{Y.~Mao and N.~Hovakimyan are with the Department of Mechanical Science and Engineering, University of Illinois at Urbana--Champaign, Urbana, IL 61801, USA (e-mail: \{ybmao, nhovakim\}@illinois.edu).}
\thanks{T.~Abdelzaher is with the Department of Computer Science, University of Illinois at Urbana-Champaign, Urbana, IL 61801, USA (e-mail: zaher@illinois.edu).}
\thanks{This work was supported in part by DOD HQ00342110002, DARPA W911NF-17-C-0099, AFOSR FA9550-15-1-0518, and NSF CNS-1932529.}}

\maketitle

\begin{abstract}
In this paper, we propose a pubic opinion model with incorporation of asymmetric cognitive bias: confirmation bias and negativity bias. We then investigate the generic modeling guidance of capturing asymmetric confirmation bias and negativity bias. A numerical examples is provided to demonstrate the correctness of
asymmetric cognitive bias model.
\end{abstract}
\IEEEpeerreviewmaketitle

\section{Introduction}

While opinion evolution models have always been an active research area, recently with the wide use of social media \cite{xu2020paradox}, in conjunction with  automated news  generation with the help of artificial intelligence technologies \cite{giridhar2019social}, it has gained a vital importance in studying misinformation spread and polarization. In this regard, confirmation bias plays a key role. Confirmation bias broadly refers to cognitive bias towards favoring information sources that affirm existing opinion \cite{nickerson1998confirmation}. It is well understood that confirmation bias helps create ``echo chambers" within networks, in which misinformation and polarization thrive, see e.g., \cite{lazer2018science,kappes2020confirmation}. Recently, Abdelzaher et al. in \cite{abdelzaher2020paradox} and Xu et al. in \cite{xu2020paradox} reveal the significant influence of consumer preferences for outlying content on opinion polarization in the  modern era of information overload. Meanwhile, Lamberson and Stuart in \cite{lamberson2018model} suggest that negative information, which is far away from expectations, is more ``outlying.'' Motivated by these discoveries, the negativity bias, which refers to a tendency to be more attentive and/or responsive to a unit of negative information than to a unit of positive information \cite{lamberson2018model}, should not be ignored in the study of information spread in social networks or public opinion evolution.

In recent few years, the Hegselmann-Krause model \cite{hegselmann2002opinion,del2017modeling,xu2020paradox} and the like-minded social influence \cite{mao2019impact,mao2020social,mao2020inference} are widely employed to capture confirmation bias. However, the considered models therein cannot fully capture the asymmetric bias, which hinders their applications in many real social problems where humans hold asymmetric confirmation bias. Motivated by the problems, we first propose a pubic opinion evolution model which explicitly takes asymmetric confirmation bias and negativity bias into account. We then investigate the generic modeling of asymmetric cognitive bias.

\section{Preliminaries}
\subsection{Notation}
The social system is composed of $n$ individuals in a social network. The interaction among individuals is modeled by a digraph $\mathfrak{G} = (\mathbb{V}, \mathbb{E})$, where $\mathbb{V}$ = $\left\{\mathrm{v}_{1}, \ldots,  \mathrm{v}_{n}\right\}$ is the set of vertices representing the individuals, and $\mathbb{E} \subseteq  \mathbb{V} \times \mathbb{V}$ is the set of edges of the digraph $\mathfrak{G}$  representing the influence structure.

\subsection{Pubic Opinion Evolution Model}
We propose the following opinion evolution model (adopted from \cite{mao2019impact}) with asymmetric cognitive bias: confirmation bias and negativity bias.
\begin{align}
{x_i}(k + 1) &= {\alpha_{i}}(k){s_i}  + \sum\limits_{j \in \mathbb{V}} c_{ij}(x(k))x_{j}(k), ~~~i \in \mathbb{V}. \label{kka}
\end{align}
Here we clarify the notations and variables:
\begin{itemize}
  \item ${x_i}\!\left( {k} \right) \in [-1,1]$ is individual $\mathrm{v}_{i}$'s opinion, ${s_i} \in [-1,1]$ is her subconscious bias, which is based on inherent personal characteristics (e.g., socio-economic conditions where the individual grew up and/or lives in) \cite{friedkin1990social}.
  \item The state-dependent influence weight $c_{ij}(x(k)) \geq 0$ is proposed to capture individual's conjunctive confirmation bias and negativity bias, which is written as
 \begin{align}
 c_{ij}(x(k)) &= \left( {1 - \beta_{i} } \right)\overline{c}( {{{\bar{x}}_i}(k),{x_{j}}(k)}) \nonumber\\
  &\hspace{1.80cm}+ \beta_{i} \underline{c}\left( {{x_i}(k),{x_{j}}(k)} \right), ~i \!\in\! \mathbb{V}\label{sdw2b}
  \end{align}
  where $\beta_{i} \in [0,1]$, $\underline{c}\left( {{x_i}(k),{x_{j}}(k)} \right)$ is used to describe confirmation bias, while $\overline{c}\left( {{{\bar{x}}_i}(k),{x_{j}}(k)} \right)$  describes negativity bias with ${{\bar x}_i}( k )$ denoting individual $\mathrm{v}_{i}$'s sensed expectation from her neighbors, defined as the mean of her neighbors' opinions, i.e.,
\begin{align}
{{\bar x}_i}( k ) \triangleq \frac{1}{{\sum\limits_{l \in \mathbb{V}} { {{w_{il}}}} }}\sum\limits_{j \in \mathbb{V}} {{{w_{ij}}}{x_j}( k )}. \label{sdw2}
\end{align}
\item  $\alpha_{i}(k) \geq 0$  is the ``resistance parameter'' of individual $\mathrm{v}_{\mathrm{i}}$ on her subconscious bias. To guarantee $x_{i}(k) \in [-1, 1]$ for $\forall k \in \mathbb{N}$ and $\forall i \in \mathbb{V}$, it is determined in the sufficient and necessary condition:
\begin{align}
{\alpha_{i}}(k) + \sum\limits_{j \in \mathbb{V}}{{c}_{ij}(x(k))x_{j}(k)}  = 1, ~~\forall i  \in \mathbb{V}.\label{sdw3}
\end{align}
\end{itemize}

\section{Asymmetric Cognitive Bias Modeling}
To simplify the presentation without loss of generality, we refer ${x_i}(k)$, ${\bar{x}_i}(k)$ and ${x_j}(k)$ to ${x_i}$, ${\bar{x}_i}$ and ${x_j}$, respectively, in this section.

\subsection{Literature Review}
We first use the examples of confirmation bias to present the modeling issues of symmetric bias. In recent few years, the Hegselmann-Krause model \cite{hegselmann2002opinion}, i.e.,
\begin{align}
\left\{ \begin{array}{l}
\!\!\!x\left( {k + 1} \right) = A\left( {x\left( k \right)} \right)x\left( k \right).\\
\!\!\!{\left[ {A(x(k))} \right]_{i,j}} = \left\{ \begin{array}{l}
 \!\!> 0,~{\underline{\varepsilon _i}} \le {x_i}(k) - {x_j}(k) < {\overline{\varepsilon _i}}\\
 \!\!= 0, ~\text{otherwise}
\end{array} \right.
\end{array} \right. \label{conf1gg}
\end{align}
and the state-dependent influence weights, i.e.,
\begin{align}
\underline{c}({x_i},{x_{j}}) = {\beta _i} - {\gamma _i}\left| {{x_i} - x_{j}} \right|,~~{\beta _i} \ge {\gamma _i} \ge 0\label{conf2gg}
\end{align}
are widely used in \cite{del2017modeling,xu2020paradox,mao2019impact,mao2020social,mao2020inference} to capture confirmation bias. However, the model \eqref{conf1gg} with asymmetric level of confidence (i.e., $|{\underline{\varepsilon _i}}| \neq {\overline{\varepsilon _i}}$) cannot fully capture the asymmetric bias while the model \eqref{conf2gg} can only capture the symmetric bias, which however hinder their applications in many real social problems where humans hold asymmetric confirmation bias. For example, we suppose the topic being discussed is ``\emph{COVID-19 Is a Hoax}", then $-1$ and $1$ correspond to individual $\mathrm{v}_{i}$ completely opposing and supporting the claim, respectively. As a consequence, $x_{j} \in [-1,0)$ means the opinion of supporting position $-1$, while $x_{j} \in (0,1]$  means the opinion of supporting position 1. We suppose that the opinions forwarded by two individuals are $x_{j} = -0.3$ and $x_{h} = 0.5$, and an individual $\mathrm{v}_{i}$'s opinion is $x_{i} = 0.1$. By that logic, and according to confirmation bias, $\mathrm{v}_{i}$ should place more weight on the opinion $x_{h} = 0.5$ than the opinion $x_{j} = -0.3$, since both $x_{h} = 0.5$ and $x_{i} = 0.1$ are supporting position 1 and the difference is their supporting degree. Yet $\left| {{x_i} - {x_j}} \right| = \left| {{x_i} - {x_h}} \right|  = 0.4$, which according to \eqref{conf2gg} and \eqref{conf1gg}, respectively, implies $\underline{c}({x_i},{x_j}) = \underline{c}({x_i},{x_h})$ and ${\left[ {A(x)} \right]_{i,j}} = {\left[ {A( x)} \right]_{i,h}}$ if $0.4 < \min\{{\overline{\varepsilon _i}}, -{\underline{\varepsilon _i}}\}$. We thus conclude that the more realistic \emph{asymmetric} confirmation bias is not captured by both the models \eqref{conf2gg} and \eqref{conf1gg}.

Building on DeGroot model \cite{degroot1974reaching}, Dandekar et al. in \cite{dandekar2013biased} proposed the opinion polarization dynamics with biased assimilation. We now examine if the model can capture the interested asymmetric confirmation bias in a simple scenario as considered in \cite{dandekar2013biased}, where the social network consists of only two individuals: $\mathrm{v}_{i}$ and $\mathrm{v}_{j}$. The proposed opinion dynamics in this scenario is ${x_i}\left( {t \!+\! 1} \right) \!=\! \frac{{{w_{ii}}{x_i}\left( t \right) + {{\left( {{x_i}\left( t \right)} \right)}^{{b_i}}}{w_{ij}}{x_j}\left( t \right)}}{{{w_{ii}} + {{\left( {{x_i}\left( t \right)} \right)}^{{b_i}}}{w_{ij}}{x_j}\left( t \right) + {{\left( {1 - {x_i}\left( t \right)} \right)}^{{b_i}}}{w_{ij}}\left( {1 - {x_j}\left( t \right)} \right)}}$, where $x_{i}(t), x_{j}(t) \in [0,1]$. If individual $\mathrm{v}_{i}$'s opinion at time $t$ is neutral, i.e., $x_{i}(t) \!=\! 0.5$. In light of the model, we have ${x_i}\left( {t \!+\! 1} \right) \!=\! \frac{{0.5{w_{ii}} + {{\left( {0.5} \right)}^{{b_i}}}{w_{ij}}{x_j}\left( t \right)}}{{{w_{ii}} + {{\left( {0.5} \right)}^{{b_i}}}{w_{ij}}}}$, which indicates that regardless of individual $\mathrm{v}_{j}$'s opinion $x_{j}(t)$, individual $\mathrm{v}_{i}$ puts the same influence weight on $x_{j}(t)$ at time $t+1$. Therefore, we can conclude the proposed polarization dynamics cannot fully capture the symmetric confirmation bias in the simplified scenario.

\subsection{Asymmetric Cognitive Bias Conditions}
The modeling challenge moving forward is \emph{How to capture the asymmetric cognitive bias?} In this section, we provide the generic modeling conditions to address the challenge.

\subsubsection{Asymmetric Confirmation Bias} It is well understood
\begin{itemize}
  \item confirmation bias happens when a person gives more weight to evidence that confirms their beliefs and undervalues evidence that could disprove it \cite{CCBB},
  \item both polarization and homogeneity are the results of the conjugate effect of confirmation bias and social influence \cite{del2017modeling,del2016spreading},
\end{itemize}
motivated by which, to capture asymmetric confirmation bias, we require the influence weights $\underline{c}(x_{i},x_{j}) \geq 0$ in \eqref{sdw2b} to satisfy
\begin{subequations}
\begin{align}
&\underline{c}(x_{i},x_{j}) > \underline{c}( {x_{i},{x_{d}}}), \nonumber\\
&\hspace{1.10cm}\text{if}~|x_{j} - x_{i} | < | {{x_{d}} - x_{i}}|, ~x_{j} \cdot x_{d} > 0,\nonumber\\
&\hspace{1.50cm}\text{or}~| {x_{j} - {x_{i}}} | = | {{{x_{d}}} - {x_{i}}}|, ~{x_{i}} \cdot x_{j} > {x_{i}} \cdot {x_{d}}, \nonumber\\
&\hspace{1.50cm}\text{or}~| {x_{j} - {x_{i}}} | = \zeta(x_{j},x_{d},x_{i}) | {{{x_{d}}} - {x_{i}}}|, ~x_{i} \cdot x_{j} < 0, \nonumber\\
&\hspace{2.00cm}x_{i} \cdot x_{d} > 0 ~\text{and}~0 < \zeta(x_{j},~x_{d},x_{i}) < 1, \label{confim3}\\
&\underline{c}( {0,x_{j}}) = \underline{c}( {0,{x_{d}}}), \hspace{0.14cm}~\text{if}~x_{j} = -{x_{d}}. \label{confim4}
\end{align}\label{confim}
\end{subequations}
\!\!\!How the proposed condition \eqref{confim} can capture the asymmetric confirmation bias are explained in the following remarks.
\begin{rem}
The condition $x_{j} \cdot x_{d} > 0$ included in the first item of the condition \eqref{confim3} means $x_{j} > 0$ \& $x_{d} > 0$, or $x_{j} < 0$ \& $x_{d} < 0$. In light of the expression, the first item indicates that for the two sensed opinions that are both supporting the position -1 or 1, the individual will put larger influence weight on the closer opinion with hers.
\end{rem}
\begin{rem} We note that the second item in the condition \eqref{confim3} includes the case: $\underline{c}( {x_{i},{u}_{d}}) > \underline{c}( {x_{i},{x_{d}}})$, if $| {x_{j} - x_{i}} | = | {{{x_{d}}} - x_{i}}|$, $x_{j} \cdot x_{i} > 0$ and ${{x_{d}}} \cdot x_{i} < 0$. If $x_{i} > 0$, we have implying  $x_{j} > 0$ and ${x_{d}} < 0$. We here conclude that in this case, although the two opinions $x_{j}$ and $x_{d}$ has the same distance with individual opinion $x_{i}$, i.e., $| {x_{j} - x_{i}} | = | {{{x_{d}}} - x_{i}}|$, individual puts larger influence weight on $u_{d}$, i.e., $\underline{c}( {x_{i},x_{j}}) > \underline{c}( {x_{i},{x_{d}}})$, since both of them support the position $1$, and vice verse as  $x_{i} < 0$. Therefore, the second item in the condition \eqref{confim3} captures the asymmetric confirmation bias when  $x_{j} \cdot x_{i} > 0$ and ${{x_{d}}} \cdot x_{i} < 0$.
\end{rem}
\begin{rem}
The second item in the condition \eqref{confim3} also includes the case: $\underline{c}( {x_{i},x_{j}}) > \underline{c}( {x_{i},{x_{d}}})$, if $| {x_{j} - x_{i}} | = | {{{x_{d}}} - x_{i}}|$ and $x_{j} \cdot x_{i} > {{x_{d}}} \cdot x_{j} > 0$. Taking $x_{i} < 0$ as an example, we have $x_{j} < {x_{d}} < 0$, by which this case implies that although $x_{j}$ and ${x_{d}}$ has the same distance with the opinion $x_{i}$ and all of them support the position $-1$, the individual $\mathrm{v}_{i}$ puts larger influence weight on $x_{j}$ since $x_{i}$ and $x_{j}$ are closer to the supporting position of $-1$ than ${x_{d}}$.
\end{rem}
\begin{rem}
Taking $x_{i} > 0$ as an example and considering $0 < \zeta(x_{j},x_{d},x_{i}) < 1$, the third item in the condition \eqref{confim3} implies that it is possible that $\underline{c}(x_{i},x_{j}) > \underline{c}( {x_{i},{x_{d}}})$, if $x_{i} < 0$, $x_{d} < 0$ and $x_{j} > 0$. This case means although both $\mathrm{v}_{i}$ and $\mathrm{v}_{d}$ support the position $-1$, while  $\mathrm{v}_{j}$ supports the position $1$, the individual puts larger influence weight on $x_{j}$ than $x_{d}$ when the ratio of their supporting-degree differences is larger than a threshold, i.e, $\frac{{\left| {x_{d} - x_{i}} \right|}}{{\left| {x_{j} - x_{i}} \right|}} \geq \frac{1}{\zeta(x_{j},x_{d},x_{i})} > 1$.
\end{rem}
\begin{rem} The condition \eqref{confim4} means that if individual's opinion is neutral, i.e., $x_{i} = 0$, she will put identical influences on her sensed opinions that have the same distance with hers.
\end{rem}

\subsubsection{Asymmetric Negativity Bias}
In this subsection, we present the conditions pertaining to capturing asymmetric negativity bias. Lamberson and Soroka in \cite{lamberson2018model} revealed that
\begin{itemize}
  \item negative information, which is far away from expectations, is more ``outlying'' (which motivates the sensed expectation $\bar{x}_{i}(k)$ in \eqref{sdw2b} and \eqref{sdw2}),
  \item the negativity bias refers to a tendency that is more attentive and/or responsive to a unit of negative information than to a unit of positive information.
\end{itemize}
Motivated by the discoveries and inspired by \eqref{confim}, to capture asymmetric negativity bias, we require the influence weights $\overline{c}\left( {{\bar{x}_i},x_{j}} \right) \geq 0$ in \eqref{sdw2b} to satisfy
\begin{subequations}
\begin{align}
&\overline{c}( {{\bar{x}_i},x_{j}}) > \overline{c}( {{\bar{x}_i},{x_{d}}}), \nonumber\\
&\hspace{1.00cm}\text{if}~| x_{j} - \bar{x}_i | > | {{x_{d}} - \bar{x}_i}|,~x_{j} \cdot x_{d} > 0,\nonumber\\
&\hspace{1.30cm}\text{or}~| {x_{j} - \bar{x}_i} | = | {{{x_{d}}} - \bar{x}_i}|, ~\bar{x}_i \cdot x_{j} < \bar{x}_i \cdot {x_{d}},\nonumber\\
&\hspace{1.30cm}\text{or}~| {{x_{d}} - \bar{x}_i} | \leq \breve{\zeta}(x_{j},x_{d},\bar{x}_i) | {{x_{j}} - \bar{x}_i}|, ~\bar{x}_i \cdot x_{j} > 0, \nonumber\\
&\hspace{2.00cm}\bar{x}_i \cdot x_{d} < 0 ~\text{and}~0 < \breve{\zeta}(x_{j},x_{d},\bar{x}_i) < 1,\label{nega2}\\
&\overline{c}( {0,x_{j}}) = \overline{c}( {0,{x_{d}}}), \hspace{0.38cm}\text{if}~x_{j} = -{x_{d}}.\label{nega3}
\end{align}\label{nega}
\end{subequations}

\subsection{Asymmetric Cognitive Bias Modeling Guidance}
In this paper, we construct $\underline{c}\left( {{x_i},x_{j}} \right)$ and $\overline{c}( {{\bar{x}_i},x_{j}})$ to have the following general forms:
\begin{align}
\underline{c}({x_i},x_{j}) &= {\underline{g_i}}( {{\underline{f_i}}({x_i}) - {\underline{f_i}}(x_{j})}) \geq 0,\label{conf2}\\
\overline{c}({\bar{x}_i},x_{j}) &= {\overline{g_i}}( {{\overline{f_i}}({\bar{x}_i}) - {\overline{f_i}}(x_{j})}) \geq 0.\label{conf3}
\end{align}

We next present the sufficient and necessary conditions of the models \eqref{conf2} and \eqref{conf3} on satisfying \eqref{confim} and \eqref{nega}, respectively, which will work as a guidance of modeling the asymmetric confirmation bias and negativity bias.
\begin{thm}
The influence weight $\underline{c}\left( {{x_i},x_{j}} \right)$ given in \eqref{conf2} satisfies \eqref{confim} if and only if
\begin{subequations}
\begin{align}
&{\underline{g_i}}( {{\underline{f_i}}(x_i) \!-\! {\underline{f_i}}x_{j}})~\text{is strictly decreasing w.r.t. the distance}\nonumber\\
&\hspace{5.35cm}|{{\underline{f_i}}( {{x_i}}) \!-\! {\underline{f_i}}( {x_{j}})} |, \label{confcc1}\\
&{\underline{f_i}}(x_{j})~\text{is strictly increasing w.r.t. $x_{j}$}, \label{confcc1ab}\\
&{\underline{f_i}}({{x_i}}) \!<\! \frac{{{\underline{f_i}}( {x_{j}}) \!+\! {\underline{f_i}}( {{{x}_d}})}}{2}, \!~\text{if}\!~| {x_{j} \!-\! {{x}_i}} | \!=\! | {{{x}_d} \!-\! {{x}_i}}|, x_{j} \!>\! {{x}_d}    \nonumber\\
&\hspace{6.24cm}\text{and}\!~{x_i} \!>\! 0, \label{confcc1ac}\\
&{\underline{f_i}}( {{x_i}}) \!>\! \frac{{{\underline{f_i}}( {{x_{j}}}) \!+\! {\underline{f_i}}( {{{x}_d}})}}{2}, \!~\text{if}\!~| {x_{j} \!-\! {{x}_i}} | \!=\! | {{{x}_d} \!-\! {{x}_i}}|, x_{j} \!<\! {{x}_d}    \nonumber\\
&\hspace{6.22cm}\text{and}\!~{x_i} \!<\! 0, \label{confcc1ad}\\
&\underline{{f_i}}(0) \!=\! \frac{{{\underline{f_i}}(x_{j}) \!+\! {\underline{f_i}}({ - x_{j}})}}{2}. \label{confcc1ae}
\end{align}\label{confcc}
\end{subequations}\label{thk1}
\end{thm}

\begin{IEEEproof} We first prove the condition \eqref{confcc} is a sufficient condition.
\subsubsection*{\underline{Sufficient Condition}} Without loss of generality, we let $x_{i} \geq x_{j} > 0$. It follows from \eqref{confcc1ab} that $| {{\underline{f_i}}(x_{i}) - {\underline{f_i}}( {{x_{j}}})} | = {{\underline{f_i}}(x_{i}) - {\underline{f_i}}( {{x_{j}}})}$. If $u_{d}$ decreases to $x_{d} > 0$, we then have $| {{\underline{f_i}}(x_{i}) - {\underline{f_i}}( {x_{j}})} | < | {{\underline{f_i}}(x_{i}) - {\underline{f_i}}( {{x_d}})} |$ and $|x_{i} - x_{j}| < |x_{i} - x_{d}|$. Considering \eqref{confcc1} and \eqref{conf2}, we have $\underline{c}(x_{i},x_{j}) > \underline{c}( {x_{i},{x_{d}}})$. If $u_{d}$ can increase to $x_{d}$ such that $x_{d} - x_{i}$ $>$ $x_{i} - x_{j} \geq 0$, we obtain from \eqref{confcc1ab} that $| {{\underline{f_i}}(x_{i}) - {\underline{f_i}}( {{x_{j}}})} | < | {{\underline{f_i}}(x_{i}) - {\underline{f_i}}( {{x_d}})} |$ and $|x_{i} - x_{j}| < |x_{i} - x_{d}|$. Considering \eqref{confcc1} and \eqref{conf2}, we then have $\underline{c}(x_{i},x_{j}) > \underline{c}( {x_{i},{x_{d}}})$. We thus conclude that the conjunctive conditions \eqref{confcc1} and \eqref{conf2} imply that
\begin{align}
\!\!\!\!\underline{c}(x_{i},x_{j}) \!>\! \underline{c}( {x_{i},\!{x_{j}}}), \!\!~\text{if}\!~|x_{d} \!-\! x_{i}| \!>\! |x_{i} \!-\!x_{j}|, x_{j} \!>\! 0, x_{d} \!>\! 0. \label{ppk1}
\end{align}
In the case of $0 > x_{i} \geq x_{j}$, following the same steps to derive \eqref{ppk1}, we have
\begin{align}
\!\!\!\!\underline{c}(x_{i},x_{j}) \!>\! \underline{c}( {x_{i},\!{x_{d}}}), \!\!~\text{if}\!~|x_{d} \!-\! x_{i}| \!>\! |x_{i} \!-\! x_{j}|, x_{j} \!<\! 0, x_{d} \!<\! 0. \label{ppk2}
\end{align}
The results \eqref{ppk1} and \eqref{ppk2} indicate that the conjunctive conditions \eqref{confcc1} and \eqref{confcc1ab} result in the first item in \eqref{confim3}.

We now consider the condition
\begin{align}
| {x_{j} - {{x}_i}} | = | {{{x}_d} - {{x}_i}}|~\text{and}~x_{j}{x_i} > {{x}_d}{x_i}. \label{cmk3as}
\end{align}
If $x_i > 0$, \eqref{cmk3as} implies that $x_{j} - {x_i} = {x_i} - {{x}_d} > 0$ and $x_{j} > {{x}_d}$, which follows from \eqref{confcc1ac} that $| {{\underline{f_i}}( {{x_{j}}}) - {\underline{f_i}}( {{x_i}})} | = {\underline{f_i}}( {{x_{j}}}) - {\underline{f_i}}( {{x_i}})$ $<$ $| {{\underline{f_i}}( {{x_i}} ) - {\underline{f_i}}( {{{x}_d}})}| = {\underline{f_i}}( {{x_i}}) - {\underline{f_i}}( {{{x}_d}} )$. We then can obtain from \eqref{confcc1} and \eqref{conf2} that \begin{align}
\!\!\!\!\underline{c}(x_{i},x_{j}) \!\!>\!\! \underline{c}( {x_{i},\!{x_{d}}}), \!\!~\text{if}\!~|x_{d} \!-\! x_{i}| \!\!=\!\! |x_{i} \!-\! x_{j}|,  x_{j} \!\!>\! {x_{d}}, x_{i} \!\!>\! 0. \label{cmk3as1}
\end{align}
If $x_i < 0$, following the same steps to derive \eqref{cmk3as1}, we have
\begin{align}
\!\!\!\!\underline{c}(x_{i},x_{j}) \!\!>\!\! \underline{c}( {x_{i},\!{x_{d}}}), \!\!~\text{if}\!~|x_{d} \!-\! x_{i}| \!\!=\!\! |x_{i} \!-\! x_{j}|,  x_{j} \!\!<\! {x_{d}}, x_{i} \!\!<\! 0. \label{cmk3as2}
\end{align}
The results \eqref{cmk3as1} and \eqref{cmk3as2} means that the conjunctive conditions \eqref{confcc1}-\eqref{confcc1ad} result in the second item in \eqref{confim3}.

Let us consider the condition
\begin{align}
x_{i}\cdot \hat{u}_d < 0 ~\text{and}~ x_{i} \cdot x_d > 0. \label{cmk3as3}
\end{align}
If $x_{i} > 0$, the condition \eqref{cmk3as3} implies that $x_j < 0$ and $x_d > 0$. Without loss of generality, we let $x_{i} < x_d - x_{i}$. In the light of \eqref{cmk3as1} and \eqref{ppk1}, we thus have
\begin{align}
\underline{c}(x_{i},x_{j}) &< \underline{c}( {x_{i},{x_{d}}}), ~~\text{if}~|x_{i} - x_{j}| = |x_{i} - x_d|\nonumber \\
\underline{c}(x_{i},0) &> \underline{c}( {x_{i},{x_{d}}}), ~~\text{if}~ 0 < x_{i} < x_d - x_{i},\nonumber
\end{align}
which indicates that there exist $x_{j} < 0$ such that $|x_{i} - x_d| > |x_{i} - x_{j}|$ and $\underline{c}(x_{i},x_{j}) > \underline{c}( {x_{i},{x_{d}}})$. Therefore, we conclude there exists an $x_{j}$ such that
\begin{align}
&\underline{c}(x_{i},x_{j}) > \underline{c}( {x_{i},{x_{d}}}), ~~\text{if}~|x_{i} - x_{j}| < |x_{i} - x_d|, x_{i} > 0, \nonumber\\
&\hspace{5.5cm}x_{j} < 0, x_d > 0. \label{akm1}
\end{align}
If $x_{i} < 0$,  following the same steps to derive \eqref{akm1}, we have conclude that there exists an $x_{j}$ such that
\begin{align}
&\underline{c}(x_{i},x_{j}) > \underline{c}( {x_{i},{x_{d}}}), ~~\text{if}~|x_{i} - x_{j}| < |x_{i} - x_d|, x_{i} < 0, \nonumber\\
&\hspace{5.5cm}x_{j} > 0, x_d < 0. \label{akm2}
\end{align}
The results \eqref{akm1} and \eqref{akm1} means that the conjunctive conditions \eqref{confcc1}-\eqref{confcc1ad} result in the third item in \eqref{confim3}.

\subsubsection*{\underline{Necessary Condition}} Given the form \eqref{conf2}, it is straightforward to verify the condition \eqref{confim4} is equivalent to \eqref{confcc1ae}. For the rest of proof, we consider contradiction, i.e, assuming \eqref{confcc1}--\eqref{confcc1ad} do not hold, the condition \eqref{confim3} does not hold as well. We assume that \eqref{confcc} does not hold, i.e., $\underline{g_i}( {{\underline{f_i}}(x_i) - {\underline{f_i}}(x_{j})})$ is nondecreasing w.r.t.$|{{\underline{f_i}}( {{x_i}}) - {\underline{f_i}}( {{x_{j}}})} |$. We let $x_{i} \geq x_{j} > 0$, and thus have $| {{\underline{f_i}}(x_{i}) - {\underline{f_i}}( {{x_{j}}})} | = {{\underline{f_i}}(x_{i}) - {\underline{f_i}}( {{x_{j}}})}$. If $x_{j}$ decreases to $x_{d} > 0$, we then have $| {{\underline{f_i}}(x_{i}) - {\underline{f_i}}( {{x_{j}}})} | < | {{\underline{f_i}}(x_{i}) - {\underline{f_i}}( {{x_d}})} |$ and $|x_{i} - x_{j}| < |x_{i} - x_{d}|$. Considering \eqref{confcc1} and \eqref{conf2}, we have
\begin{align}
\underline{c}(x_{i},x_{j}) \leq \underline{c}( {x_{i},{x_{d}}}). \label{pkmq1}
\end{align}
We now consider the case that $\underline{g_i}( {{\underline{f_i}}(x_i) - {\underline{f_i}}(x_{j})})$ is strictly non-decreasing w.r.t.$|{{\underline{f_i}}( {{x_i}}) - {\underline{f_i}}( {{x_{j}}})} |$ and ${\underline{f_i}}(\cdot)$ is strictly non-increasing w.r.t. $c$. Let us set $0 < x_d < x_{j} < x_{i}$. We thus have $|x_{i} - x_d| > |x_{i} - x_{j}|$, $|{{\underline{f_i}}( {{x_i}}) - {\underline{f_i}}( {{x_d}})} | \geq |{{\underline{f_i}}( {{x_i}}) - {\underline{f_i}}( {{x_{j}}})} |$
and \eqref{pkmq1}. Following the same analysis method, we can conclude that if the conditions \eqref{confcc1}--\eqref{confcc1ad} do not hold, the \eqref{confim3} does not hold as well.
\end{IEEEproof}

\begin{thm}
$\overline{c}\left( {{\bar{x}_i}(k),{x_{j}}(k)} \right)$ given in \eqref{conf3} satisfies \eqref{nega} if and only if
  \begin{subequations}
\begin{align}
&{\overline{g_i}}( {{\overline{f_i}}({{\bar{x}_i}}) \!-\! {\overline{f_i}}( {{x_{j}}})})~\text{is strictly increasing w.r.t. the distance}\nonumber\\
&\hspace{5.25cm} |{{\overline{f_i}}( {{\bar{x}_i}}) \!-\! {\overline{f_i}}( {{x_{j}}})} |, \label{confcc1k}\\
&{\overline{f_i}}(x_{j})~\text{is strictly increasing w.r.t. $x_{j}$}, \label{confcc3k}\\
&{\overline{f_i}}( {{\bar{x}_i}}) \!<\! \frac{{{\overline{f_i}}( {x_{j}}) \!+\! {\overline{f_i}}( {{{x}_d}})}}{2}, \!~\text{if}\!~| {x_{j} \!-\! {\bar{x}_i}} | \!=\! | {{{x}_d} \!-\! {\bar{x}_i}}|, x_{j} \!>\! {{x}_d}    \nonumber\\
&\hspace{6.24cm}\text{and}\!~{\bar{x}_i} \!>\! 0, \label{confcc4k}\\
&{\overline{f_i}}( {{\bar{x}_i}}) \!>\! \frac{{{\overline{f_i}}( {{x_{j}}}) \!+\! {\overline{f_i}}( {{{x}_d}})}}{2}, \!~\text{if}\!~| {x_{j} \!-\! {\bar{x}_i}} | \!=\! | {{{x}_d} \!-\! {{x}_i}}|, x_{j} \!<\! {{x}_d}    \nonumber\\
&\hspace{6.22cm}\text{and}\!~{\bar{x}_i} \!<\! 0, \label{confcc5k}\\
&{\overline{f_i}}(0) \!=\! \frac{{{\overline{f_i}}(x_{j}) \!+\! {\overline{f_i}}( { - x_{j}})}}{2}. \label{confcc2k}
\end{align}\label{confcck}
\end{subequations}\label{thk2}
\end{thm}

\begin{IEEEproof}
The proof steps completely follow those of Theorem \ref{thk1}, it is thus omitted.
\end{IEEEproof}

\begin{figure}
\centering
\includegraphics[scale=0.31]{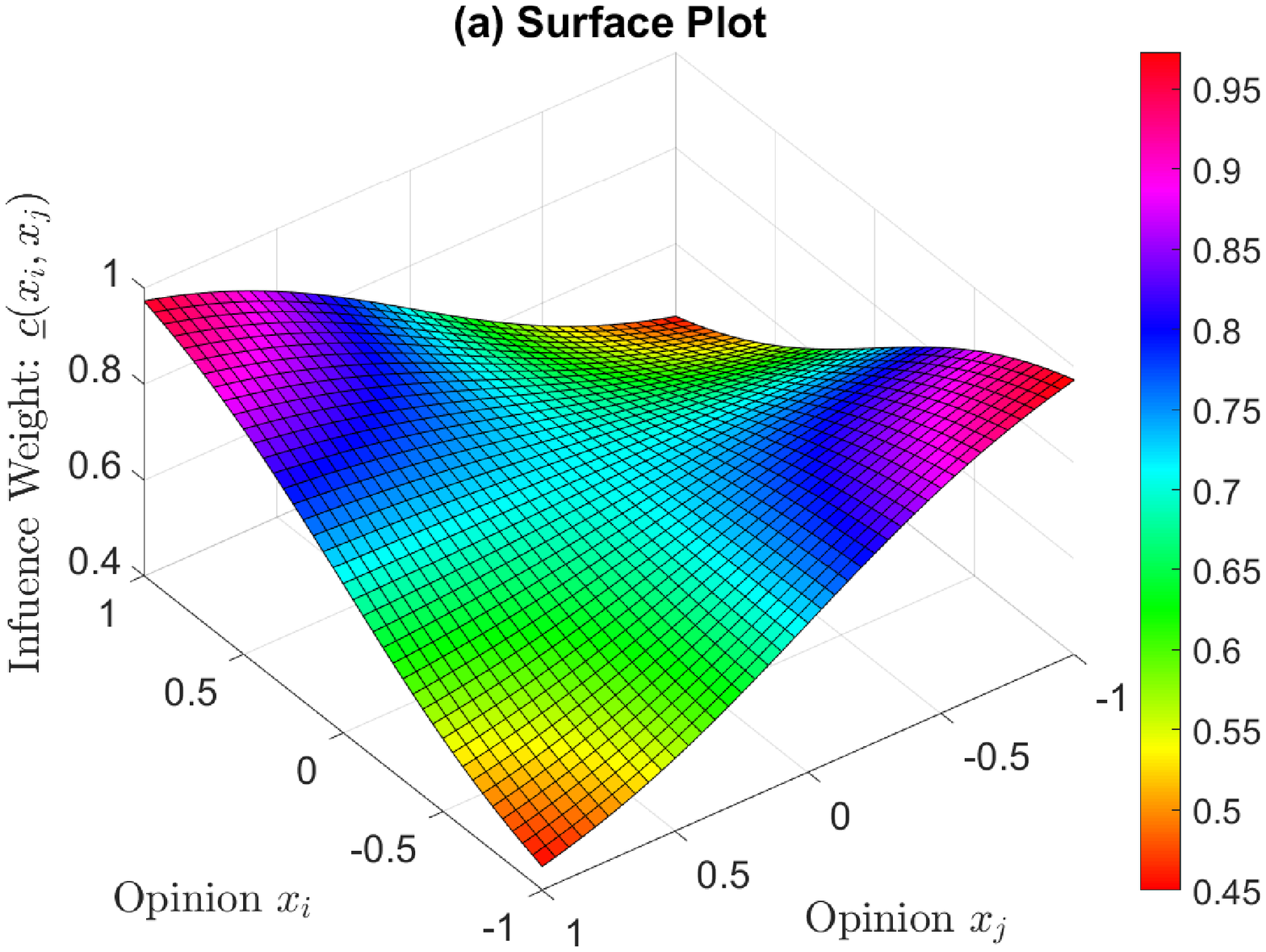}\\
\includegraphics[scale=0.31]{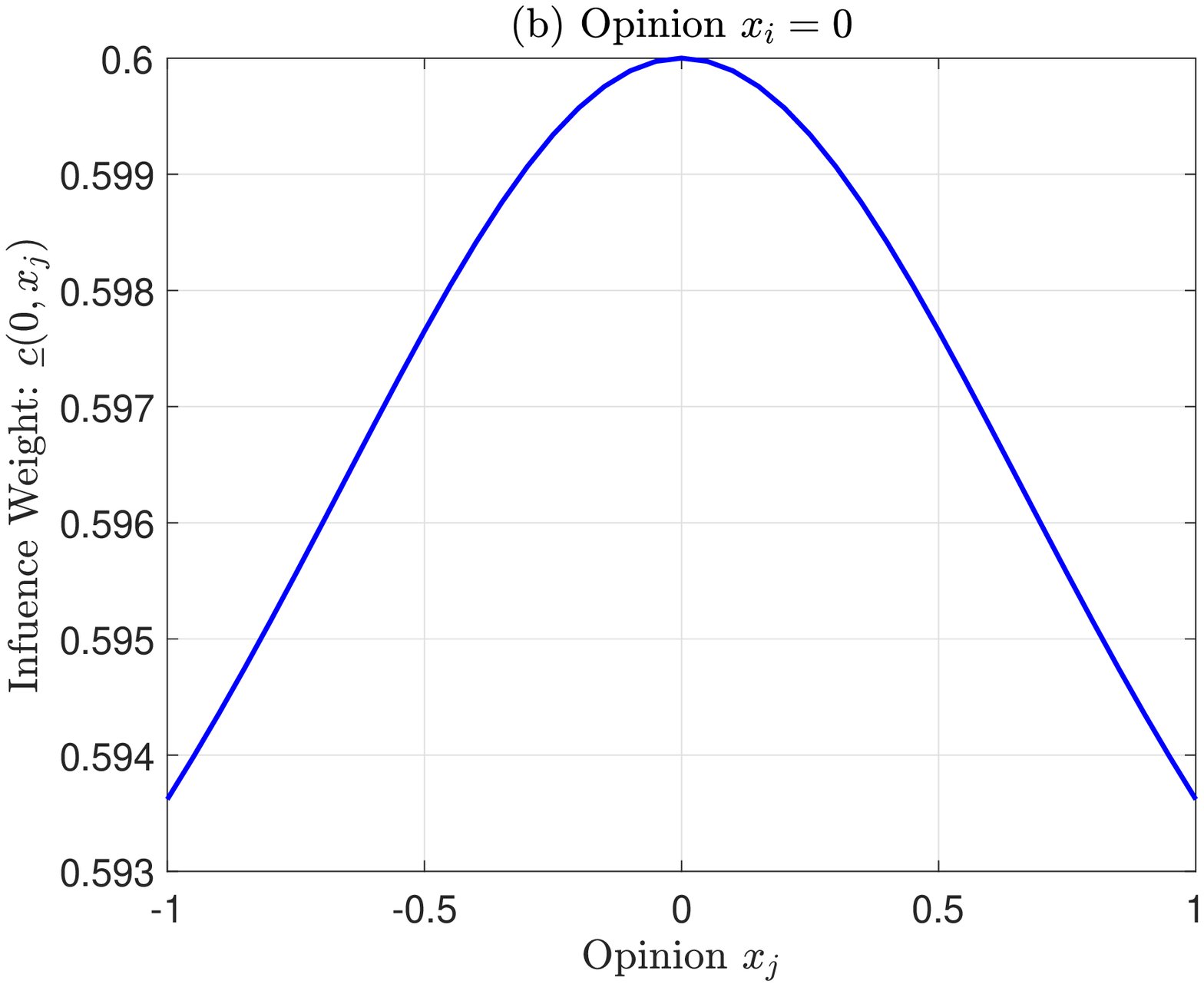}\\
\includegraphics[scale=0.31]{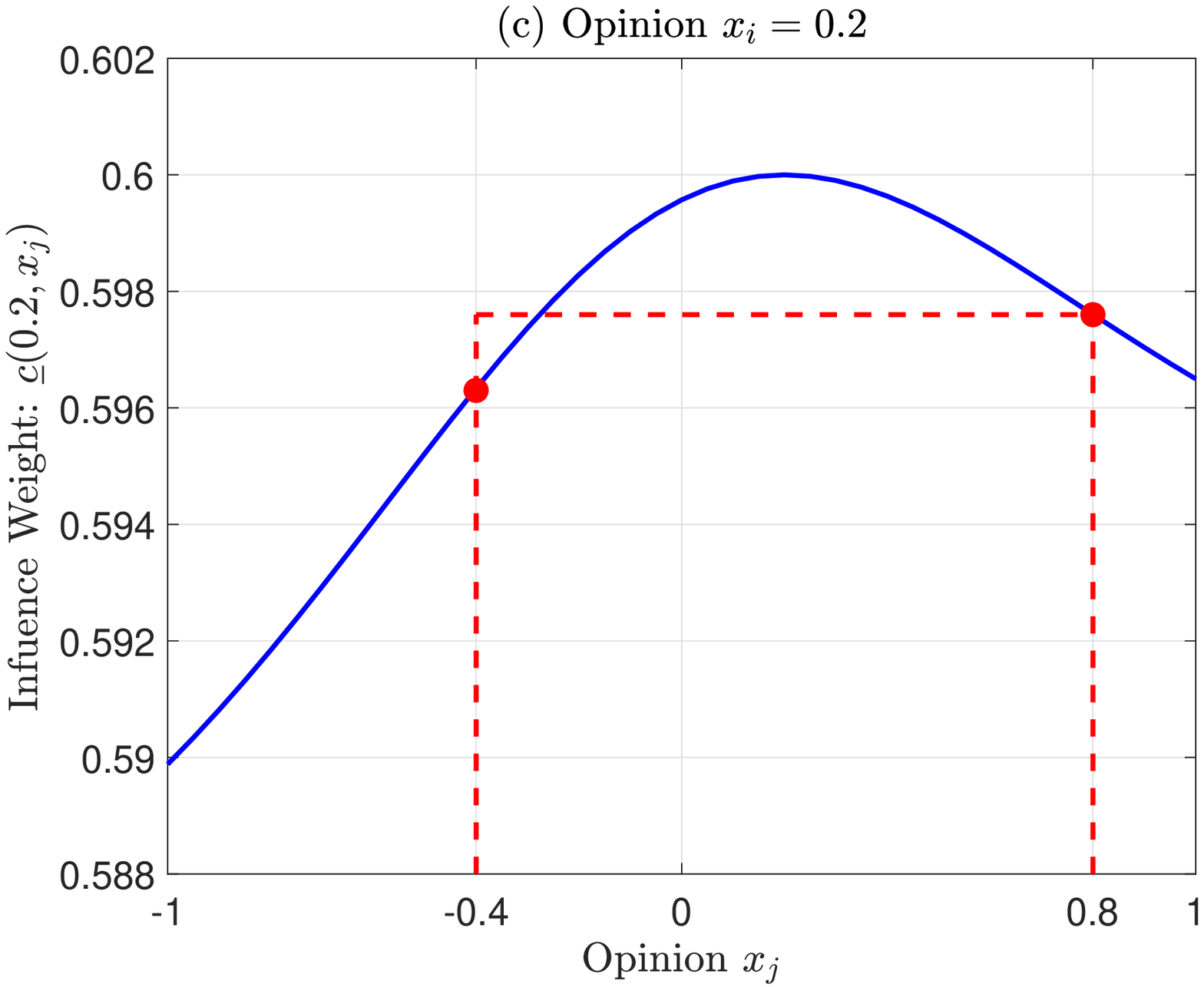}\\
\includegraphics[scale=0.31]{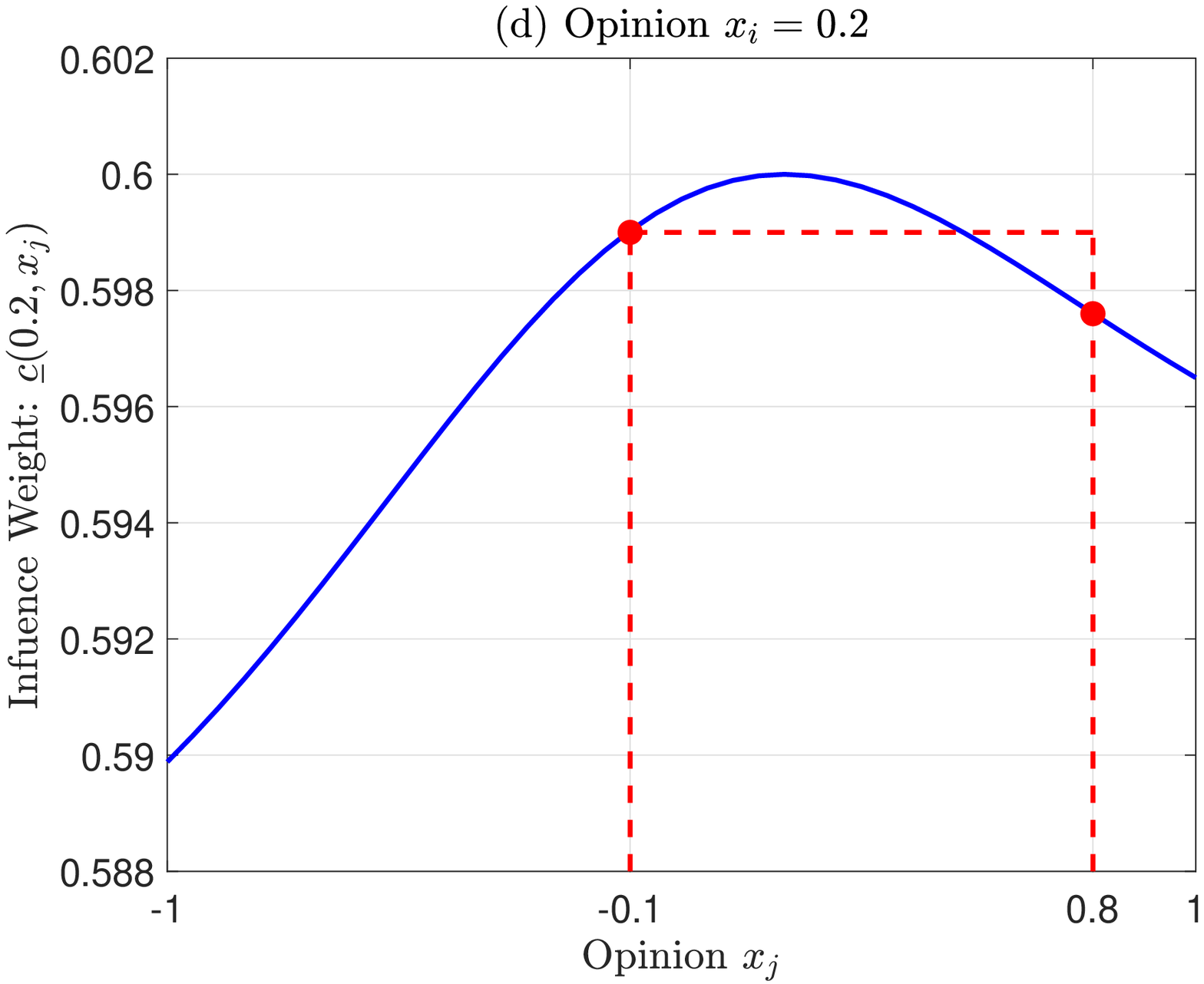}\\
\includegraphics[scale=0.31]{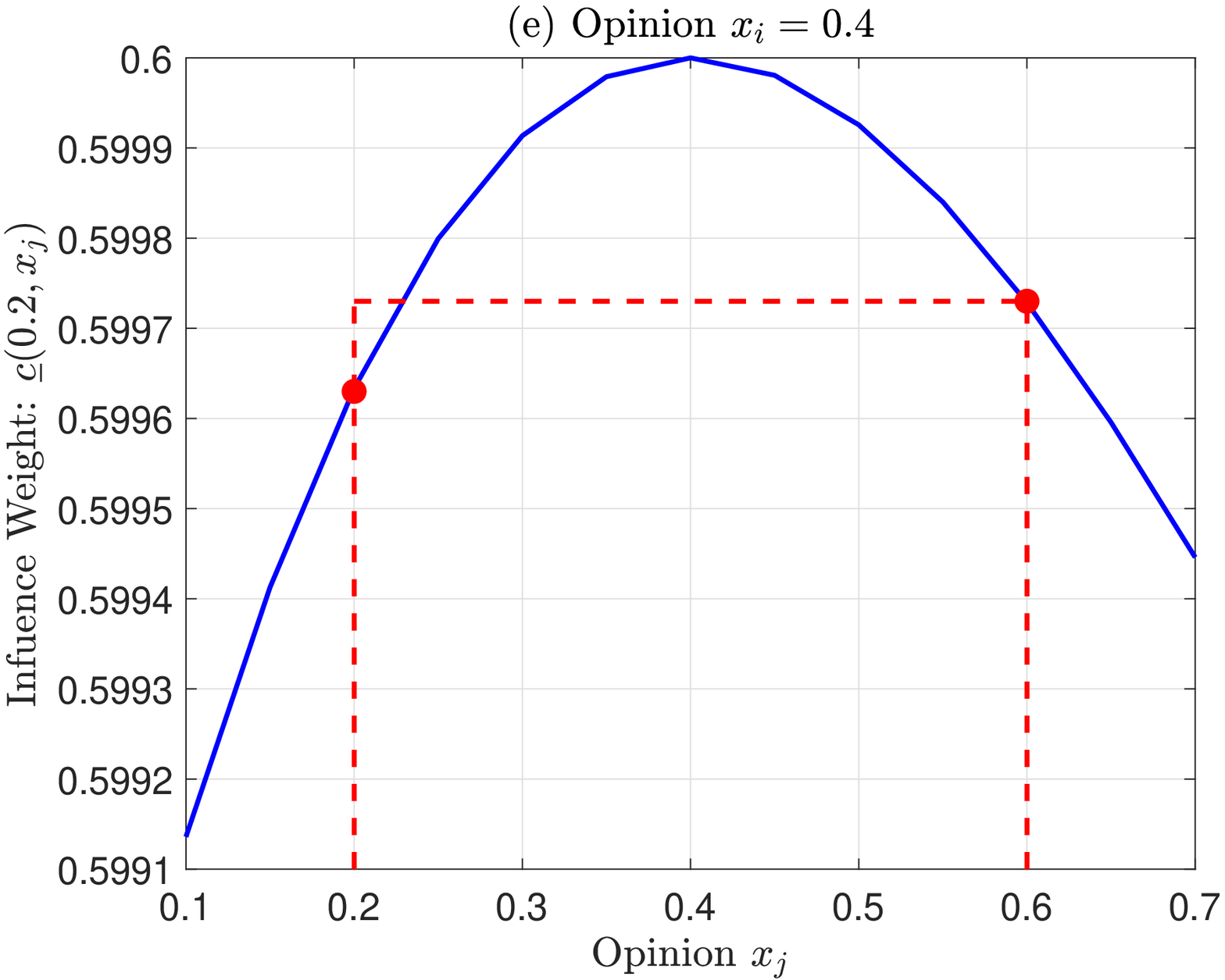}
\caption{Influence weight function  $\underline{c}(x_{i},x_{j}) = 0.6 - 0.011(\tanh(x_{i}) - \tanh(x_{j}))^2$ captures the asymmetric confirmation bias.}
\label{spcc}
\end{figure}

\subsection{Numerical Example}

Theorems \ref{thk1} and \ref{thk2} provides the guides to construct the models of asymmetric confirmation bias and negativity bias, respectively. The models are not unique. Take the confirmation bias as an example, its models include $\underline{c}( {{x_i},{x_{j}}}) = {\chi _i} - {\gamma _i}( {\tanh({{x_i}}) - \tanh ( {{x_{j}}})})^{2}$, $\underline{c}( {{x_i},{x_{j}}}) = {\chi _i} - {\gamma _i}| {x_i^3 - x_{j}^3} |$, with $\gamma _i > 0$,  and among many others. We now use $\underline{c}(x_{i},x_{j}) = 0.6 - 0.011(\tanh(x_{i}) - \tanh(x_{j}))^2$ as one numerical example to demonstrate its checkable properties, observing its values shown in Figure \ref{spcc}, we can verify that
\begin{itemize}
  \item If individual $\mathrm{v}_{i}$'s opinion is neutral, i.e., $x_{i} = 0$, she puts the identical influence weights on the opinions that have same distance with hers.
  \item If individual $\mathrm{v}_{i}$ supports the position $1$ with $x_{i} = 0.2$, for the two opinions $0.8$  and $-0.4$ that have the same distance with $x_{i} = 0.2$, she puts larger influence weight on $0.8$, since $0.8$ and $0.2$ are in the same supporting domain.
    \item If individual $\mathrm{v}_{i}$ supports the position $1$ with $x_{i} = 0.2$, for the two opinions $0.8$  and $-0.1$, she puts larger influence weight on $-0.1$, since although $0.2$ and $0.8$ are in the same supporting domain while $-0.1$ is in the opposing domain, $0.8$ has much larger distance with  $x_{i} = 0.2$  compared with $-0.1$.
  \item If individual $\mathrm{v}_{i}$ supports the position $1$ with $x_{i} = 0.4$, for the two opinions $0.6$  and $0.2$ that have the same distance with $x_{i} = 0.4$, she puts larger influence weight on $0.6$, since all of them are in the same supporting domain but both $0.6$ and $0.4$ stick more to the supporting domain compared with $0.2$.
\end{itemize}

\bibliographystyle{IEEEtran}
\bibliography{ref}

\end{document}